# Tailoring Magnetic Properties of CoFeB Films via Tungsten Buffer and Capping Layers

L. Saravanan[1*], Nanhe Kumar Gupta[2], Carlos Garcia[1], and Sujeet Chaudhary[2]

[1]*Departamento de Física, Universidad Técnica Federico Santa María, Valparaíso 2390123, Chile*
[2]*Thin Film Laboratory, Department of Physics, Indian Institute of Technology Delhi, New Delhi 110016, India*

[*]saravanan.lakshmanan@usm.cl

## Abstract

Controlling the interface between W and CoFeB-based buffer or capping layers at an appropriate temperature is essential for modifying the strength of magnetic anisotropy. In this work, we systematically explore the impact of W buffer and capping layers on the structural, topological, and magnetic anisotropy properties of W (5 nm)/CoFeB(10 nm) and CoFeB(10 nm)/W(3 nm) bilayers sputtered at room temperature (RT) and annealed at an optimal annealing temperature ($T_A$) of 400°C. Our findings demonstrate that the bilayer films' uniaxial magnetic anisotropy (UMA) with out-of-plane coercivity ($H_{c\perp}$) is highly influenced by the W buffer, capping layers, and $T_A$. Specifically, the $H_{c\perp}$ of the CoFeB layer with the buffer and capping layers annealed at 400°C samples exceed several times the coercivity of those unannealed. CoFeB buffered with W and annealed at 400°C shows larger $H_{c\perp}$, two-fold UMA, and higher in-plane UMA energy density ($K_{eff}$) than the CoFeB/W bilayers, which can be attributed to the W buffer layer inducing the crystallization of CoFeB during annealing. The W buffer, capping layers, and the $T_A$ for W and CoFeB-based bilayer samples significantly alter the surface morphology, grain sizes, and surface roughness. The XRD analysis reveals nano-crystallites embedded in the larger grains of the 400°C annealed samples. Hence, this work offers a promising approach to achieving high thermal stability of UMA in W and CoFeB-based spintronic applications.

# 1. INTRODUCTION

There is a broad interest in studying ultrathin films and nanostructures with perpendicular magnetic anisotropy (PMA) and large $H_{c\perp}$ for their application in novel spintronic devices. Recent studies have revealed that ultrathin films composed of heavy metal (HM)/ferromagnet (FM) layers can significantly elucidate spin-orbit effects such as Spin-Orbit Torque (SOT) [1] and Dzyaloshinskii-Moriya Interaction (DMI) [2]. The DMI in ultrathin films with specific magnetic anisotropy is critical to enhancing topologically protected spin structures such as magnetic skyrmions [3,4]. The SOT mechanism enables switching the magnetization in a ferromagnetic film by applying a current through an adjacent heavy metal layer or topological insulators [5,6].

The specific properties and interfaces of thin films are heavily influenced by their preparation methods and growth conditions. The properties include lattice mismatches between adjacent layers, crystallographic relationships, surface roughness, surface morphology, interface characteristics, and electronic effects [7,8]. For practical applications, magnetic films must possess adjustable magnetic coercivity, domain structure, and magnetization distribution [9]. This adjustability is achieved through engineering magnetic, buffer, and capping layers, involving thermal treatments, film thickness, and magnetic film composition modifications. Magnetic coercivity values depend on the films' magnetic anisotropy energy, crystal structure, and surface morphology [10–12]. These findings have prompted a systematic investigation into the impact of heavy metal buffers and capping layers with a large Spin Hall Angle ($\theta_{SHA}$) on the magnetic anisotropy of CoFeB films deposited on the suitable substrates. Recent studies have contributed to optimizing the Magnetic Tunnel Junction (MTJ) structures by integrating specific heavy metals such as Ta, Pt, Hf, Mo, Ru, and W as buffer and/or capping layers [13–18]. Specifically, research has shown that a CoFeB/MgO framework with W buffer and capping layers exhibits superior annealing stability compared to a CoFeB/MgO framework with Ta buffer and capping layers [17]. This suggests that incorporating specific heavy metals can enhance the performance and reliability of MTJ structures and offer potential improvements in various applications such as magnetic memory devices and other spintronic technologies.

PMA and/or $H_{c\perp}$ properties at lower thicknesses of the magnetic layer are crucial for ultra-thin film data storage applications [19–21]. Additionally, modern perpendicular MTJs (p-MTJs) must withstand industrial standard processing temperatures of 400°C necessitating satisfactory thermal stability of the stacks [22,23]. B. Sun *et al.* [24] first reported a significant enhancement in the $H_{c\perp}$ of the CoFeB layer in Si//CoFeB/Ag structures with a thickness of 172 nm, which is attributed to magnetocrystalline anisotropy, Additionally, the effect on $H_{c\perp}$ of thickness, buffer and capping layers of Ta or W in CoFeB films have been previously investigated [25–27]. A significantly high value of $H_{c\perp}$ of 1250 Oe was reported for W/CoFeB/W at 400°C trilayer films [27]. However, the development of CoFeB films with W buffer and capping layers alone and a comprehensive study of their structural, surface morphological and magnetic properties have not yet been reported. From both fundamental and technological perspectives, examining interface-induced $H_{c\perp}$ in designing multilayer structures with relatively thick CoFeB films is relevant to understanding the role of W capping and buffer layers in tailoring magnetic properties and optimizing hard-axis coercivity for advanced spintronic devices. Thicker CoFeB films are a model for studying fundamental magnetic properties such as spin waves, domain walls, and magnetization dynamics. Specifically, the study of 10 nm thickness CoFeB films is significant because i) it provides a high surface-to-volume ratio, ii) it leads to low power consumption in spintronic devices, iii) it enhances fast magnetization dynamics for high-speed applications, and iv) it ensures compatibility with existing semiconductor manufacturing processes.

This study reports the structural, magnetic anisotropy, and surface morphological properties of W(5 nm)/CoFeB(10 nm) and CoFeB(10 nm)/W(3 nm) bilayers annealed at an optimal temperature of 400°C. The findings highlight the critical role of tungsten buffer and capping layers in tuning uniaxial magnetic anisotropy and enhancing hard axis coercivity.

## 2. EXPERIMENTAL DETAILS

Bilayer films were fabricated on thermally oxidized (500 nm $SiO_2$) Si substrates [sizes = 1 × 1 square inch] using the pulsed DC magnetron sputtering technique at RT. After preparation, samples were immediately characterized to minimize the formation of surface impurities. Two types of heterostructures were deposited:

I. $SiO_2$//W (5 nm)/CoFeB(10 nm)
II. $SiO_2$//CoFeB(10 nm)/W(3 nm)

The base pressure in the sputtering chamber was maintained below $1.0 \times 10^{-6}$ Torr. The sputtering targets of $Co_{20}Fe_{60}B_{20}$ and W with high purity (99.99%) were used to grow the respective bilayers. The ferromagnetic CoFeB layer was deposited using a DC power of 40 W and an Ar pressure of 3.2 mTorr. The heavy metal W layer was sputtered at a DC power of 10 W and an Ar pressure of 3.2 mTorr. The sputtering rates for CoFeB and W were 0.83 Å/s and 0.26 Å/s, respectively. After deposition, the bilayers underwent *ex-situ* annealing at the optimal temperature of 400°C for 60 minutes in a vacuum environment below $4.0 \times 10^{-5}$ Torr without applying any magnetic field.

The crystal structure of the bilayer films was analyzed using Grazing Incidence X-ray Diffraction (GI-XRD) with a *PANalytical X'pert PRO* diffractometer and employing Cu $K_\alpha$ radiation ($\lambda = 1.5406$ Å). The thickness of the individual layers within the structure was determined through X-ray Reflectivity (XRR) measurements using the same diffractometer with the *X'Pert Reflectivity, v1.2a* software. The elemental composition of the CoFeB films (Co: Fe: B = 20: 60: 20) was confirmed by Energy Dispersive Spectroscopy (EDS) coupled with Scanning Electron Microscopy (SEM) *[Hitachi High Technol]*. The magnetic anisotropy measurements were conducted at RT using a home-built longitudinal Magneto-Optical Kerr Effect (L-MOKE) setup. This instrument utilized a linearly polarized intensity stabilized He-Ne laser source ($\lambda = 632$ nm) with a laser spot diameter of less than 500 µm [28]. The angular dependence of the squareness ratio ($M_r/M_{s//}$) was determined by recording MOKE M-H loops at different in-plane azimuthal orientations of the applied magnetic field. In-plane and out-of-plane DC magnetization hysteresis loops were also recorded at RT using the Vibrating Sample Magnetometer (VSM) *[Microsense EZ11 VSM]*. The surface morphology of the samples was examined using Atomic Force Microscopy (AFM) *[Asylum Research MFP3D-SA]* in tapping mode.

## 3. RESULTS AND DISCUSSION

### 3.1 Thin film thickness measurement

Fig. 1 (a)-(b) schematically illustrates the two types of fabricated heterostructures comprising layers of W and CoFeB-based sputtered on $SiO_2$ (500 nm) substrates. The CoFeB layer thicknesses are estimated from XRR fitting and aligned well with the nominal film thickness of 10 nm. To accurately determine the thickness of each layer, XRR measurements were conducted on all the as-dep. or RT bilayer films as shown in Fig. 1(c). The fitted XRR spectra closely match the observed experimental data.

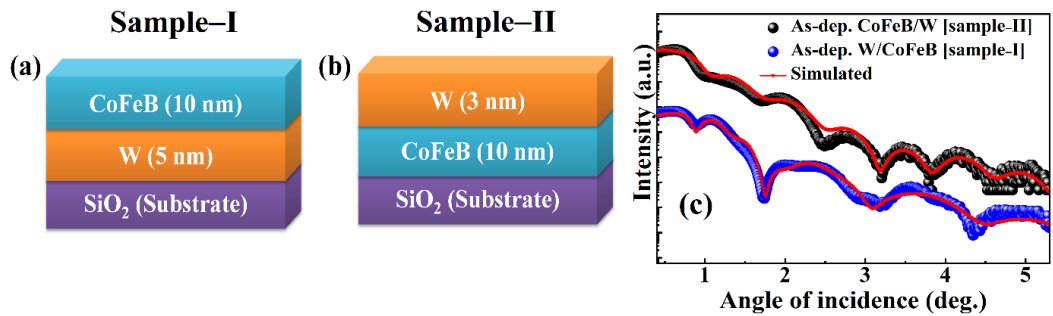

**Fig.1** Schematic representation of the two types of heterostructures investigated in our experiment: (a) W/CoFeB and (b) CoFeB/W samples [labeled as sample–I and sample–II] and (c) The observed and simulated X-ray reflectivity spectrum of as-dep. heterostructures of samples–I and II.

The persistent presence of Kiessig fringes (XRR oscillations) up to $2\theta \approx 6°$ indicates the films' high surface quality and relationship with the interface. The thickness, film density, and roughness interface of the as-dep. samples were determined by fitting experimentally obtained XRR profiles (see Table 1). The density of each layer is determined to be either very close to or slightly smaller than their respective bulk values. The measured thickness of individual layers matches the nominal thickness values. The overall interface roughness is less than 0.8 nm, indicating the excellent quality of these CoFeB films sputtered under high vacuum conditions. The analysis of the XRR profiles indicates a significant increase in interface width between CoFeB and W or Ta with crystallization [29].

**Table 1.** The fitted parameters for the thickness (t), density ($\Gamma$), and interface roughness ($\sigma_i$) of an as-dep. $SiO_2$//W(5nm)/CoFeB(10nm) and $SiO_2$//CoFeB(10nm)/W(3nm) films.

| Sample | Simulated Parameters | Layers | | | | |
| --- | --- | --- | --- | --- | --- | --- |
| | | SiO$_2$ | W | CoFeB | W | Oxide layer |
| **W/CoFeB** | Γ(g/cc) | 2.12 ± 0.03 | 18.12 ± 0.08 | 6.7 ± 0.6 | --- | 1.51 ± 0.03 |
| | t(nm) | 565 ± 14 | 5.2 ± 0.3 | 10.5 ± 0.6 | --- | 1.33 ± 0.05 |
| | σ$_i$ (nm) | 0.23 ± 0.03 | 0.41 ± 0.06 | 0.49 ± 0.06 | --- | 0.41 ± 0.03 |
| **CoFeB/W** | Γ(g/cc) | 2.764 | --- | 7.0 ± 0.5 | 16.78 ± 0.05 | 1.54 ± 0.02 |
| | t(nm) | 521 ± 12 | --- | 9.3 ± 0.6 | 2.7 ± 0.2 | 0.93 ± 0.03 |
| | σ$_i$ (nm) | 0.14 ± 0.01 | --- | 0.32 ± 0.05 | 0.31 ± 0.04 | 0.72 ± 0.06 |

## *3.2 Structural analysis*

The GI-XRD patterns were obtained from all the as-dep. CoFeB (10 nm) films indicate their amorphous nature and no distinct intensity signals corresponding to CoFeB or W films were observed (see Fig. 2). This amorphous characteristic is further supported by HR-TEM analysis conducted by M. Raju *et al.* [29]. Ref [27] shows that all the *ex-situ* annealed W(5 nm)/Co$_{20}$Fe$_{60}$B$_{20}$(10 nm)/W(3 nm) trilayer samples displayed a prominent crystalline peak corresponding to the (110) plane at $2\theta \approx 45°$ which is indicative of the CoFe crystalline phase. Additionally, Nanhe *et al.* observed that *in-situ* annealed Ta (12 nm)/Co$_{60}$Fe$_{20}$B$_{20}$(28 nm)/Ta (5 nm) trilayer samples exhibited an amorphous nature up to temperatures of ≤ 400°C. However, annealing at temperatures ≥ 415°C resulted in the transformation of CoFeB from an amorphous to a crystalline CoFe phase [29].

In our system, samples–I and II annealed at 400°C show polycrystalline bcc-CoFe, as evidenced by diffraction patterns at $2\theta = 44.9°$, attributed to the (110) plane. The crystallite size of CoFe is determined by the Debye-Scherrer formula,

$$D = \frac{K\lambda}{\beta Cos\,\theta} \quad (1)$$

where D, K, λ, and β are the average crystallite size, a constant related to the size distribution, X-ray wavelength, and full width at half maximum of the peak, respectively [30]. The average crystallite size of CoFe was determined to be ~ 18 nm and ~ 11 nm for samples–I and II at 400°C, respectively, and this result is consistent with previous reports [27].

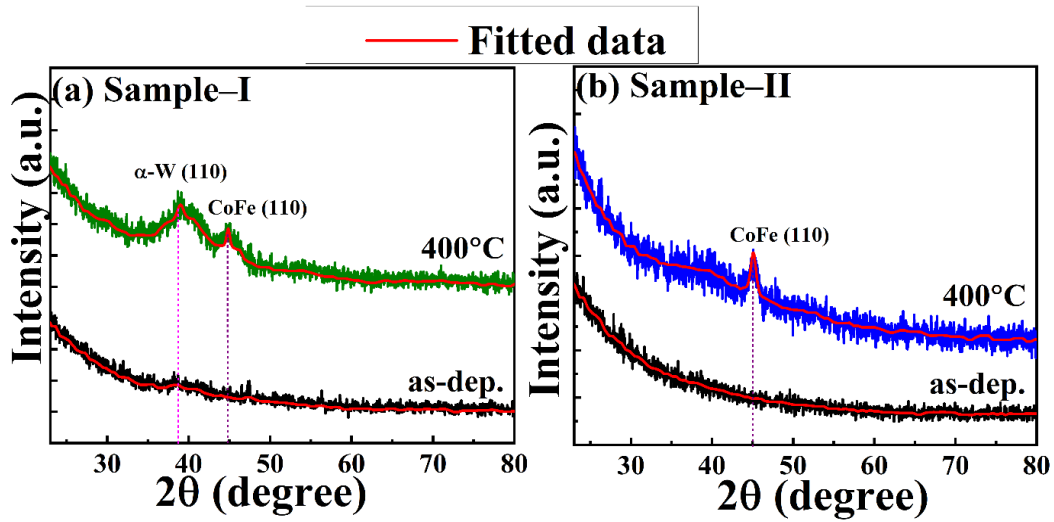

**Fig.2** GI-XRD patterns for (a) sample–I and (b) sample–II and both sputtered at RT (as-dep.) and followed by annealing at 400°C.

The W buffer and capping layers influence the crystallite size variation in CoFeB films at 400°C. The crystalline phase is more pronounced in the W/CoFeB bilayer compared to the CoFeB/W bilayer. XRD analysis also reveals the presence of the α-W phase with a crystallite size of ~9 nm in the sample–I at 400°C and indicates a stable α-W phase formed on thermally oxidized Si substrates for thicker (5 nm) W films. This result aligns with previous reports [30,31] showing that sputtered W films' phase and resistivity depend on deposition conditions, temperature and thickness.

### 3.3 Magnetic anisotropy properties
#### 3.3.1 In-plane magnetic anisotropy: L-MOKE

To study the magnetic anisotropy in W/CoFeB and CoFeB/W bilayers under various temperature conditions, we investigated the magnetization reversal of the samples using MOKE with an external magnetic field applied in both easy and hard axis directions. The in-plane angle-dependent MOKE hysteresis curves were recorded at RT from 0º to 360º with a step size 10º for both as-dep. and annealed (at 400ºC) W(5 nm)/CoFeB(10 nm) and CoFeB(10 nm)/W(3 nm) samples. The results (see Fig. 3) suggest significant magnetic anisotropy exists in the sample–I at 400ºC. The changes in in-plane coercivity depend on various factors such as defects pinning the motion of domain walls, the extent of oxidation, crystallite size distribution, the creation of voids between grains, and so on [32,33]. The magnetization reversal in CoFe films was reported previously [34].

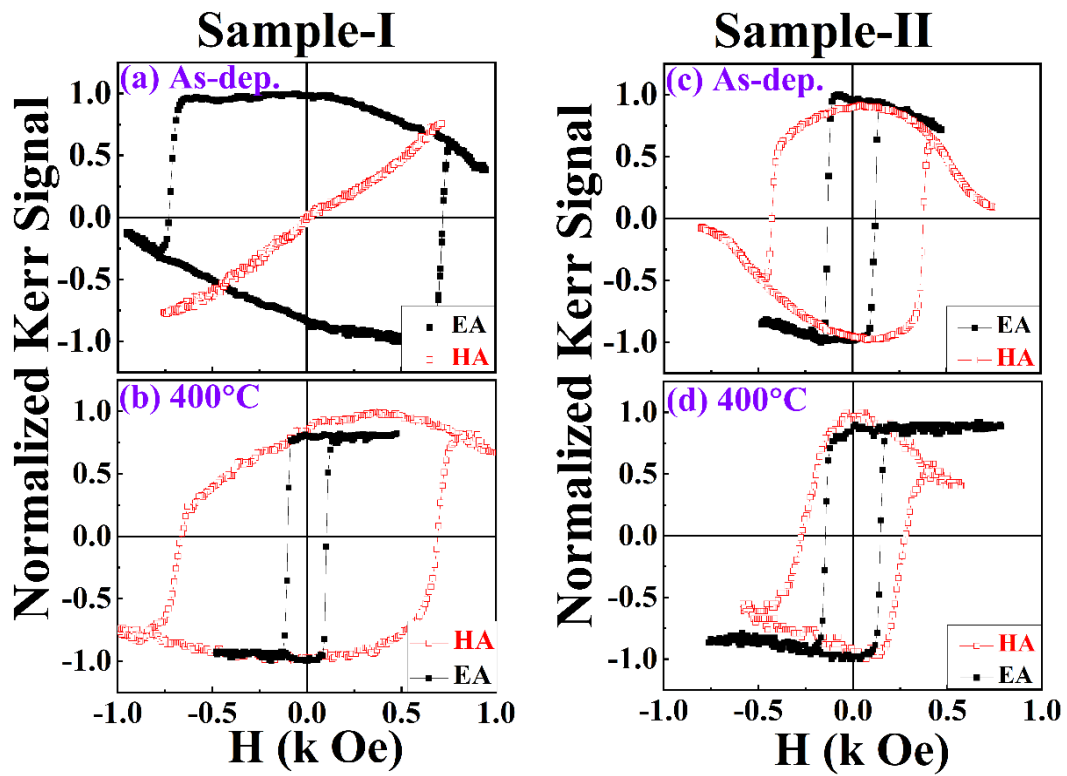

**Fig.3 (a)-(d)** MOKE M-H hysteresis loops recorded on samples–I and II in both the as-dep. and they annealed (400°C) states for two field orientations corresponding to the application of external magnetic field along the in-plane easy (black colour solid-squared data symbols) and hard axes (red colour open-squared data symbols). Lines are a guide to the eye.

The MOKE hysteresis loops corresponding to cases where the field is applied along the in-plane easy axis (*EA*) and hard axis (*HA*) of various W/CoFeB and CoFeB/W bilayers are shown in Fig. 3(a)-(d). For sample–I grown at RT, a large in-plane coercivity ($H_{c//}$) in the hysteresis loops was recorded along the easy axis [in-plane $H_{//}$ *EA*], contrasting with the absence of $H_{c//}$ observed for the hard axis [in-plane $H_{//}$ *HA*]. In the 400°C annealed sample–I, a small $H_{c//}$ of M-H loops was recorded along the easy axis, whereas a large $H_{c//}$ was obtained along the hard axis, as shown in Fig. 3(b). Moreover, a small $H_{c//}$ of hysteresis curves was again observed along the easy axis for sample–II at 400°C films (Fig. 3(d)). Previous reports [34] indicated that variations in the $H_{c//}$ value along the easy direction mainly depend on manipulating parameters such as pinning holes, sputtering bias voltage, crystallite size, deposition rates of the films, etc. Additionally, local structural properties such as alterations in the grain sizes, defects, and strain [35] also affect the $H_{c//}$ and $M_r/M_{s//}$.

Figure 4(a)-(d) shows the angular-dependent $M_r/M_{s//}$ values of samples–I and II at RT and annealed at 400ºC. This elucidates a consistent two-fold symmetry with differing shapes. From this analysis, we infer that the angular dependent $M_r/M_{s//}$ in the samples–I and II [both as-dep. and at 400°C] show well-defined strong magnetic anisotropy with two-fold symmetry. The detailed observation of two-fold anisotropy with various shapes was reported previously [27,36]. This in-plane UMA with two-fold symmetry is often observed in sputtered amorphous CoFe-based alloy systems and is believed to arise from strain effect and an anisotropic chemical disorder [37].

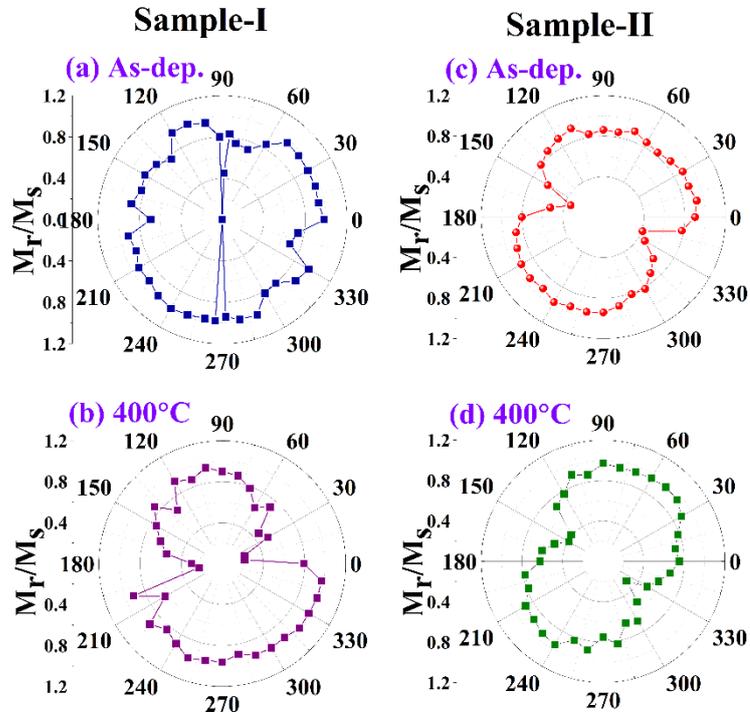

**Fig.4 (a)-(d)** Angular dependence of normalized $M_r/M_s$ as a function of the applied field orientation for the sample–I and sample–II bilayer samples at RT and after annealing at 400°C. These samples demonstrate uniaxial symmetry along both the easy and hard axes.

*3.3.2. In-plane and out-of-plane magnetic anisotropy: VSM*

We further investigated the magnetic anisotropy (in-plane and out-of-plane directions) properties of the CoFeB film with W buffer and capping layers to assess the influence of $T_A$ changes on the magnetostatic properties. The magnetic measurements were performed at RT utilizing a VSM by applying magnetic field strengths ranging from +5.0 kOe to -5.0 kOe in parallel/in-plane and perpendicular/out-of-plane orientations of the film's surface. Figs. 5(a)-(d) presents the normalized M-H loops of both as-dep. and 400°C annealed samples–I and II. All the bilayer films exhibit low coercivity along the in-plane direction, corresponding to the easy axis of magnetization. In contrast, the hysteresis curves along the out-of-plane direction show high coercivity, indicating it serves as the hard axis of magnetization. The anisotropy field ($H_k$) was estimated from the hard axis saturation field measured at the intersection of in-plane and out-of-plane hysteresis loops [38,39]. The positive and negative signs of the $H_k$ indicate the presence of PMA and in-plane magnetic anisotropy (IPA), respectively. The parallel and perpendicular M-H curves of samples–I and II at RT are displayed in Fig. 5 (a) and (c), respectively and all our bilayer samples exhibit IPA properties. The W buffer and capping layers significantly influence the strength of the $H_k$ for the CoFeB film and are evidenced by the $H_k$ values of -1.92 ± 0.08 kOe for as-dep. sample–I and -1.82 ± 0.07 kOe for as-dep. sample–II. A similar trend is observed at a $T_A$ of 400°C for sample–I and sample–II as shown in Fig. 5 (b) and (d) [$H_k$ values of -2.6 ± 0.1 kOe and -2.05 ± 0.08 kOe are observed for samples–I and II, respectively].

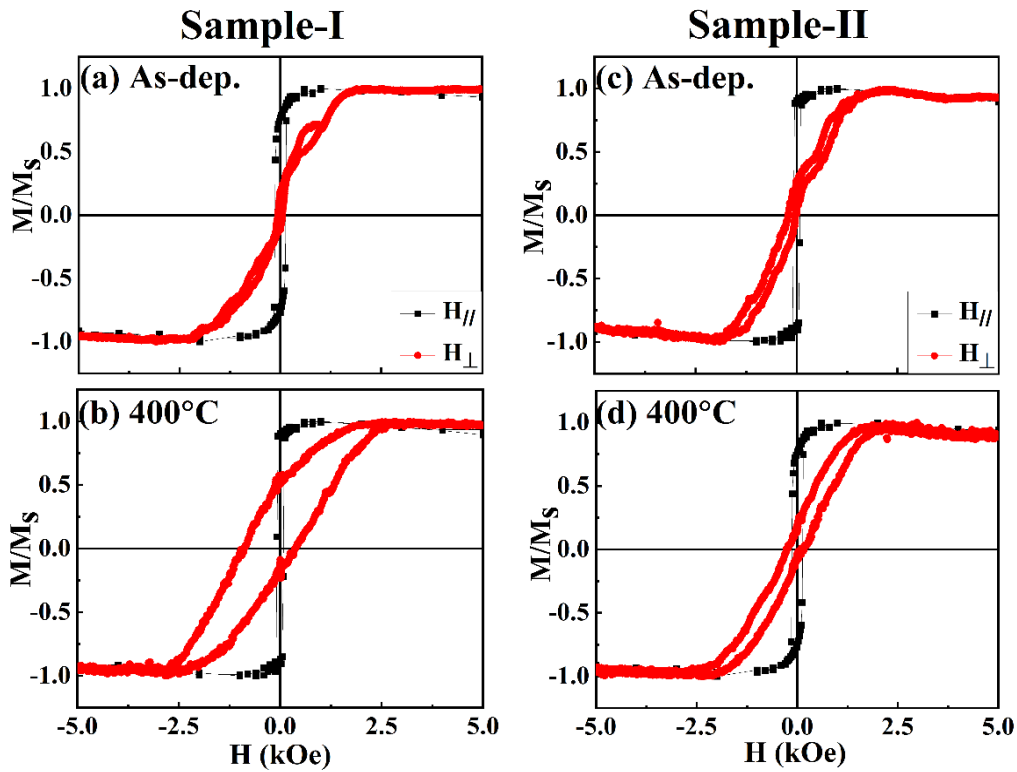

**Fig.5** (a) & (c) Hysteresis loops of sample–I and sample–II in the as-dep. state, (b) & (d) sample–I and sample–II annealed at $T_A$ = 400°C.

Annealing at an appropriate temperature can rearrange the constituent elements at the interfaces, thereby minimizing the disorders that may have formed during deposition [40]. Therefore, the optimum $T_A$ is crucial to adjusting the magnetic anisotropy in the films [37]. The $H_{c\perp}$ of the sample–I was found to be 52 ± 2 Oe and 636 ± 25 Oe for as-dep. and 400°C annealed state, respectively. Similarly, for sample–II, $H_{c\perp}$ values of 91 ± 3 Oe and 206 ± 9 Oe were determined for the as-dep. and 400°C annealed states, respectively. The $H_{c\perp}$ values of a 400°C annealed sample are several times larger than those in the as-dep. state. A larger $H_{c\perp}$ is observed for sample–I compared to sample–II at 400°C, which could be attributed to the W buffer layer promoting the crystallization of CoFeB during annealing [27,29]. Similar effects of buffer and capping layers in CoFeB/MgO multilayers at different temperatures have been recently reported [41]. Moreover, the release of boron atoms from CoFeB film and its microstructure can significantly impact the $H_{c\perp}$ and magnetic moment during heat treatment [25–27,29,42–44]. Boron atoms typically have a smaller atomic radius compared to Co, Fe, and W. This size difference results in higher mobility and faster diffusion of boron atoms relative to these elements. During annealing, the increased mobility of boron atoms induces changes in the magnetization behavior of the CoFeB film.

The crystalline phase of CoFeB in the W/CoFeB sample is more pronounced than in the CoFeB/W bilayer samples annealed at 400°C. This difference in intensity arises from the fact that nanocrystallites in the W/CoFeB sample exert tensile stress on the surrounding amorphous phase and vice versa [47,48]. Additionally, several other factors may contribute to the observed large $H_{c\perp}$ in the W/CoFeB bilayer: i) some bond orientation order resulting from interfacial interaction with the W layer [45,46], ii) structural defects and changes in compositional homogeneities due to crystallization can hinder domain wall motion and may also contribute to increased coercivity, iii) grain size and surface roughness of the films could affect $H_{c\perp}$ and perpendicular anisotropy [25,47], iv) domain nucleation, growth and annihilation, where the average sizes develop intensely and may correlate strongly with strong $H_{c\perp}$ [48]. Mainly, magnetization reversal in the out-of-plane direction of the samples is likely generated by effectively single nucleation due to strong inter-granular coupling and a phenomenon significantly induced by using the W layer [49,50]. Therefore, such out-of-plane magnetization reversal could be more strongly facilitated in the W/CoFeB bilayer than in the CoFeB/W bilayer. L-MOKE results on the angular dependence of $M_r/M_s$ for all the bilayer samples at RT and 400°C (Fig. 4) revealed a relatively stronger two-fold magnetic anisotropy in the annealed W/CoFeB bilayer than in the CoFeB/W system.

The release of boron from the CoFeB into the interface can lead to the formation of boron oxide and a well-ordered CoFe crystal structure under thermal treatment, which is one of the critical aspects primarily enhancing the $H_{c\perp}$ [25]. In consequence, the $H_{c\perp}$ in the CoFeB film is strongly influenced by the choice of buffer and capping layers [24,25,27].

The $K_{eff}$ is obtained by [51],

$$K_{eff} = \frac{M_S H_K}{2} \tag{2}$$

To understand the strength of interfacial in-plane anisotropy in CoFeB films, we investigated the dependence of $K_{eff}$ on $T_A$ for W/CoFeB and CoFeB/W bilayers. The $K_{eff}$ depends on both $K_u$ and saturation magnetization ($M_s$), as indicated in equation (3), representing the cumulative impact of multiple anisotropy contributions on $K_{eff}$,

$$K_{eff} = K_U - 2\pi M_S^2 \tag{3}$$

where,

$$K_U = \left[ K_V + \left(\frac{2K_i}{t}\right) + \left(\frac{3}{2}\lambda\sigma\right) \right] \tag{4}$$

Here, $t$ represents the thickness of the magnetic layer. Breaking down equations (3) and (4), the right-hand side comprises four distinct terms, each contributing to the overall $K_{eff}$ of the system. In equation (4), the first term corresponds to volume magnetic anisotropy ($K_v$), which is linked to the crystallographic structure of the multilayer. The second term represents interface/surface magnetic anisotropy ($K_i$), arising from the interaction between electronic *d-p* and *d-d* states at the interfaces of different layers such as HM/FM/oxide or HM/FM [52]. The third term accounts for magnetoelastic anisotropy with '$\sigma$' and '$\lambda$' denoting the stress and magnetostriction constants. This term captures how mechanical stress and strain affect the material's magnetic properties [38,53]. Lastly, the fourth term in equation (3) denotes shape anisotropy $-2\pi M_S^2$ and favors an easy-plane anisotropy. This term accounts for the demagnetization energy associated with the film's shape and influences the preferred orientation of magnetization within the sample [54,55]. According to equation (3), a negative value of $K_{eff}$ indicates a preference for the easy axis of magnetization to align parallel to the plane within the samples, suggesting that demagnetization energy dominates over PMA. Conversely, a positive value of $K_{eff}$ suggests that the easy axis of magnetization tends to align perpendicular to the plane.

As shown in Table 2, the IPA gets stronger when the films are annealed from RT to 400°C for both samples–I and II. With the increase in $T_A$, $K_{eff}$ initially exhibits a higher value and then decreases. The maximum negative $K_{eff}$ value is estimated at -8.1 ± 0.3 × $10^5$ erg/cc for sample–I annealed at 400°C. A similar trend is observed for sample–II annealed at 400°C, with the maximum negative $K_{eff}$ value of -5.8 ± 0.2 × $10^5$ erg/cc. Similar changes in the magnetic and structural properties were recorded in CoFeB film-based stacks [29,37,51].

**Table 2.** In-plane and out-of-plane $H_c$, $H_k$, and $K_{eff}$ for samples–I and II at various temperatures, especially as-dep. and 400°C.

| S.No | Sample ID | $T_A$ (°C) | Magnetic properties | | | |
|---|---|---|---|---|---|---|
| | | | $H_{c//}$ (Oe) | $H_{c\perp}$ (Oe) | $H_k$ (k Oe) | $K_{eff}$ (×$10^5$ erg/cc) |
| 1 | I | As-dep. | 134±5 | 52±2 | -1.92 ±0.08 | -5.4 ±0.2 |
| 2 | | 400 | 78±3 | 636±25 | -2.6 ±0.1 | -8.1 ±0.3 |
| 3 | II | As-dep. | 78±3 | 91±3 | -1.82 ±0.08 | -5.6 ±0.2 |
| 4 | | 400 | 132±5 | 206±9 | -2.06 ±0.08 | -5.8 ±0.2 |

*3.4 Topological analysis*

The surface morphology of samples–I and II was investigated by operating AFM in tapping mode within a scan range of 5 × 5 μm² and a scale range of 1 × 1 μm². The AFM images shown in Fig. 6 demonstrate the impact of $T_A$ on the bilayer films. The variations in surface morphology, surface roughness ($R_{rms}$), and grain size as a function of $T_A$ are evident in both samples–I and II. For the as-dep. and 400°C annealed sample–I, distinct topographies are observed: the as-dep. sample displays individual, nearly spherical grains ranging from ≈ 118 nm to ≈ 142 nm (Fig. 6(a)), while the 400ºC annealed sample shows rod-shaped grains as displayed in Fig. 6(b). In contrast, the as-dep. sample–II features spherical grains of ≈ 72 nm in size. The annealed sample–II exhibits larger agglomerated grains of ≈ 332 nm (Fig. 6(d)), which indicates that the post-annealing process led to agglomeration and further crystallization. The grain size and morphology can influence the coercivity of the films for both applied field orientations (in-plane and out-of-plane) with rod-shaped or larger grains responsible for higher coercivity. This observation aligns with earlier reports [27,56,57].

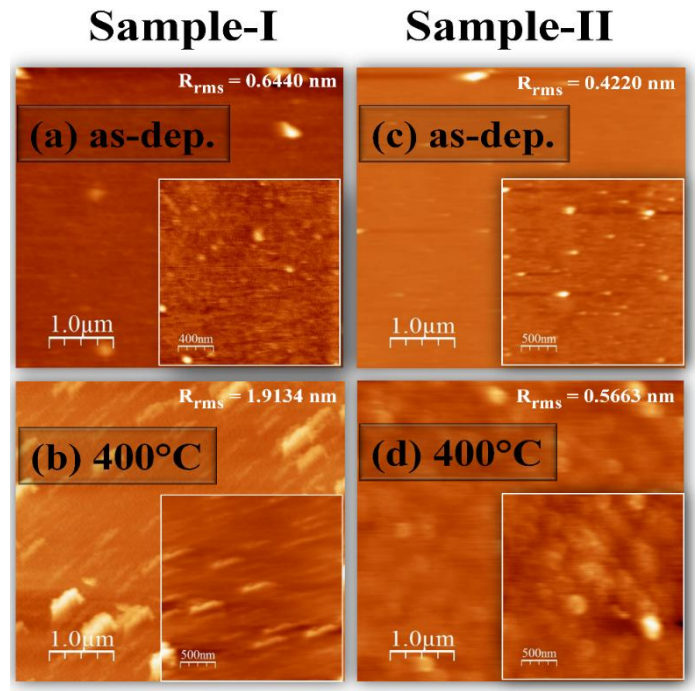

**Fig.6** AFM images captured on the surface of the bilayer films in both as-dep. and 400°C annealed states: [(a)-(b)] for sample–I and [(c)-(d)] for sample–II.

The $R_{rms}$ values for the as-dep. and 400 ºC annealed W/CoFeB films are ≈ 0.64 nm and ≈ 1.91 nm, respectively. Similarly, $R_{rms}$ values of ≈ 0.42 nm at RT/as-dep. and ≈ 0.57 nm at 400ºC were observed for the CoFeB/W samples. The lowest interface/surface roughness is observed in the as-dep. sample-II, while an abrupt increase in roughness occurs at higher $T_A$. Generally, annealing promotes surface smoothness in materials. However, an increase in $R_{rms}$ at higher annealing temperatures could also occur, as was observed in our case. This increase is often attributed to the coalescence of large-sized crystallites and/or the growth of grain size [58]. Additionally. the surface roughness and grain size can influence $H_{c\perp}$ [25,47]. On the other hand, the absence of a capping layer in the sample–I may result in slight oxidation, which could alter the grain size on the surface.

The interface and surface roughness values obtained from XRR fitting (see Table 1) and AFM analysis show a similar trend across the samples–I and –II sequences. It is worth noting that interface roughness comparisons were made exclusively for the as-dep. samples–I and –II bilayers. Additionally, an oxide layer is present in both as-dep. samples-I and-II.

We claim the presence of magnetic anisotropy in our as-dep. amorphous W/CoFeB and CoFeB/W samples based on bond-orientational anisotropy (BOA). The structural anisotropy arising from the anisotropic structure factor and pair distribution function in binary or multi-component glassy systems was first reported by Suzuki *et al.* in $Fe_{40}Ni_{40}Mo_3Si_{12}B_5$ [59]. They linked this structural feature to atomic BOA, induced by stresses, where more atomic bonds align in one direction than perpendicular in glassy and amorphous solids. Harris *et al.* later confirmed BOA in sputtered amorphous Tb-Fe films and tied it to the uneven distribution of Fe-Fe, Tb-Tb, and Fe-Tb pairs within the film plane versus perpendicular orientations, which induces magnetic anisotropy in the film [60]. Such microscopic structural differences are inevitable in thin-film systems due to growth kinetics and associated strain and stress.

The magnetization anisotropy in amorphous CoFeB films grown via a magnetron sputtering system has been attributed to BOA and influenced by variations in composition (particularly boron) and thickness [45]. Considering the comparable energy kinetics observed during the growth of our samples, we consider that the exact BOA mechanism could also occur in our as-dep. W/CoFeB and CoFeB/W samples. In contrast, the annealed samples feature fewer but larger grains. This change leads to significant enhancements in nucleation and/or domain wall pinning and a substantial increase in anisotropy, possibly resulting in greater coercivity [27,48,61,62]. Consequently, the annealed CoFeB and W-based samples with larger grains exhibit higher coercivity than the as-dep. samples with smaller grains. Specifically, in our case, the grain size in the W/CoFeB annealed at 400ºC is larger than that of CoFeB/W annealed at the same temperature. The annealing temperature of 400°C is energetically more favorable for segregating boron, resulting in a reduced boron content and altered magnetic moment in CoFeB. This observation is consistent with our XRD and VSM results.

## 4. SUMMARY AND CONCLUSION

In summary, the role of the tungsten buffer and capping layers in CoFeB film at a suitable temperature of 400°C emerges as a crucial factor in controlling the desired UMA properties. We have observed out-of-plane coercivity and UMA in as-dep. and annealed at 400°C optimal annealed W/CoFeB and CoFeB/W bilayers with a 10 nm thick CoFeB film. A large $H_{c\perp}$ of 636 ± 25 Oe with two-fold symmetry UMA is observed for the W/CoFeB bilayer annealed at 400°C. The presence of large $H_{c\perp}$ could have originated due to CoFeB crystallization, domain nucleation, and variations in microstructure during annealing. The annealing and the presence of buffer and capping layers of W considerably modify the UMA energy and create competition between $H_{c\perp}$ and in-plane UMA. A maximum in-plane $K_{eff}$ of $8.1 \pm 0.3 \times 10^5$ erg/cc is achieved for the W/CoFeB bilayer at 400°C sample. The $\alpha$-W in the W/CoFeB bilayer annealed at 400°C enhances $H_{c\perp}$ through various mechanisms, including crystal structure stability, enhancing magnetic anisotropy, exchange coupling, grain size and microstructural. This research explores a significant avenue for studying and developing the modern generation of ultra-low power storage applications, particularly in SOT-MRAM.

## ACKNOWLEDGMENTS

L. Saravanan acknowledges the funding support from the FONDECYT Postdoctorado (2022), Agencia Nacional de Investigación y Desarrollo (ANID) (grant # 3220373). C. Garcia acknowledges the financial support ANID FONDECYT/Regular 1241918 and ANID FONDEQUIP EQM140161 received. This work was also supported by the European Union's Horizon 2020 research and innovation program under the Marie Sklodowska-Curie Grant Agreement No. 101007825 (ULTIMATE-I project).

# HIGHLIGHTS

- All W/CoFeB and CoFeB/W bilayer samples exhibited magnetic anisotropy with two-fold symmetry

- A large $H_{c\perp}$ of 636 ± 25 Oe and $K_{eff}$ of 8.1 ± 0.3×10$^5$ erg/cc for in-plane UMA is observed for W/CoFeB bilayer annealed at 400°C

- High $H_{c\perp}$ could result from improved CoFeB crystallization with α-W phase formation and annealing-induced microstructure alterations

<div style="text-align:center">*****</div>